\documentclass[aps,prl,twocolumn,superscriptaddress,nofootinbib,
nobalancelastpage,showpacs,nobibnotes]{revtex4}
\usepackage{epsfig}

\begin{document}
\title
{Plasma effects on resonant fusion
}

\author{R. F. Sawyer}
\affiliation{Department of Physics, University of California at
Santa Barbara, Santa Barbara, California 93106}

\begin{abstract}
We investigate the effects of plasma interactions on
resonance-enhanced fusion rates in stars.  Starting from basic principles we derive an expression for
the fusion rate that can serve as a basis for discussion of approximation schemes.
The present state-of-the-art correction algorithms, based on the classical correlation function for the fusing particles 
and the classical energy shift for the resonant state, do not follow from this result, even as an approximation.
The results of expanding in a perturbation solution for the case of a weakly coupled plasma are somewhat
enlightening.
But at this point we are at a loss as to how to do meaningful
calculations in systems with even moderate plasma coupling strength.
Examples where this can matter are: the effect of a possible low energy $^{12}$ C +$^{12}$ C resonance on X-ray bursts from accreting neutron stars or on supernova 1A simulations;
and the calculation of the triple $\alpha$ rate in some of the more strongly coupled regions in which the process
enters, such as accretion onto a neutron star.

\end{abstract}
\maketitle

\section{1. Introduction}

Nuclear fusion rates play a central role in stellar evolution theory.  Their calculation
depends primarily on laboratory data, and the open questions that persist
have mostly to do with these measurements and their extrapolations
to very low energies. Another issue that presents itself is that of the influence
of the surrounding plasma on the fusion rates. This is a mature subject, to say the least,
for the case of non-resonant fusion, but there are some questions that still are not settled, and some of these pertain to the
special case of resonance-enhanced fusion, which is the subject of the body of this paper.

To set the context, we need to begin with a review of the way the plasma corrections work
for the ordinary case in which the energy variation of the vacuum value of [cross-section$\times$ velocity] is
mainly from the Coulomb wave-functions, rather than from a resonant peak.

A lore has developed, the ``basically classical" approach \cite{salpeter}- \cite{it3},
which follows the
following path: 

1.) calculation of an effective two particle correlation function (down to a small distance)
in a classical simulation; 

2.) defining an effective two body potential proportional to the logarithm
of this function, written in the form $V_{\rm eff}=Z^2 e^2/r+V_{\rm sc}(r)$ where $V_{\rm sc}$ is referred to as the
``screening potential"; 

3.) in the potential $V_{\rm eff}$
calculating the tunneling through the barrier to determine
the correlation function at zero separation $K(r=0)$. This correlator then is assumed to embody \underline{all} of the
effects of the plasma on the fusion rate. 

Implementing the above, we consider fusing particles A and B, with respective number densities, $n_a$
and $n_b$. We write the rate, $w$, at which one specified nucleus of type $B$ undergoes fusion as,
\begin{eqnarray}
w=n_a  \langle \sigma v \rangle {K(0) \over K_0(0)} \,.
\label{old}
\end{eqnarray}
and we consider all of the energy dependence of the vacuum cross-section $\sigma$ to come from
the Coulomb potential.
The factor $K_0(0)$ is the correlator in the absence of the plasma, and serves to cancel
the Coulomb factor that comes in $\sigma$. We should emphasize that everything
in the present paper pertains to cases in the thermo-nuclear regime, rather than in the pychnonuclear
regime. 

Turning to the resonant case, in the absence of plasma
the contribution of a narrow resonance to the fusion rate is, 
\begin{eqnarray}
 w_0=n_a \zeta e^{-E_r/T} \exp \Bigr( -\pi e^2 Z^2 \sqrt { {\mu \over E_r}} ~~\Bigr)  \,,
 \label{ans1}
\end{eqnarray}
where $E_r$ is the energy and $\Gamma^{(0)}_r$  is what the partial width of the resonance in the channel of the fusing nuclei
would be in the absence  of the Coulomb force between the decay products. The final factor in (\ref{ans1})
is the exponential factor in the Coulomb wave-function; and $\mu$ is the reduced mass of the fusing particles. The
parameter $\zeta$ is easily calculated from the development of section 2: it does not enter the ratios that are here addressed.

This paper discusses approaches to incorporating plasma effects into the rates
for the resonant case.
These plasma effects can be important in applications. Refs. \cite{cussons} and \cite{cooper} suggest that the existence a resonant state in the $^{12}$C+$^{12}C$ system in the vicinity
of $E_r=$1.5 MeV in energy might explain
features of X ray bursts from accreting neutron stars. Refs. \cite{cussons} and \cite{itoh} discuss the screening corrections
to resonant rates, motivated by this possibility.  Following these references, we choose for our examples a density of
$\rho=5\times {\rm10^9 gc^{-3}}$ for a pure $^{12}{\rm C}$ plasma, and a temperature region $4<T_8<9$.

For purposes of exposition
we assume that the screening potential $V_{\rm sc}$ is nearly constant in the classically inaccessible region for the two
particle system with resonant energy, and make the replacement $V_{\rm sc}(r)\approx V_{\rm sc}(0)$ in this region.
In the above application this is a good approximation\footnote{It is
easy to go beyond this simplification, using the mechanics presented in ref. \cite{itoh}.}. We write $V_{\rm sc}(0)=-\gamma T$ and take the dimensionless parameter
$\gamma$ from ref. \cite{itoh},
\begin{eqnarray}
 \gamma=2.293 \Gamma^{0.25}+1.053 \Gamma-0.5551 \log (\Gamma)-2.35\,,
 \label{gamma}
\end{eqnarray}
where $\Gamma$ is the usual plasma coupling. Under our sample conditions of composition and density we have
 $\Gamma=.62 /T_9$.
A first guess as to the form for the screened rate formula could be,
  \begin{eqnarray}
 w_{\rm scr}=n_a \zeta  e^{-E_r/T} \exp \Bigr( {-\pi e^2 Z^2 \sqrt  \mu \over \sqrt{E_r+\gamma T}}~\Bigr),
\nonumber\\
\,
 \label{ans2}
\end{eqnarray}
where the penetration factor has been altered by subtracting the screening potential at the origin, $-\gamma T$, from the 
energy in the incoming state. In normal fusion, where there is an integral over energy and the other energy dependent factors are the phase
space and the statistical factor $\exp[ E/T]$, a simple shift of integration variable by $-\gamma T$
gives the usual enhancement factor $\exp \beta \gamma $ in rate, but this does not apply to the resonance case, it appears. 

In refs. \cite{cussons} and \cite{itoh},  
the result (\ref{ans2}) is extended by including a plasma-induced resonance-energy
shift $\Delta E_{r}$ given by making the replacement, $E_r\rightarrow E_r+\Delta_{E_{r}}$, and leading to,
\begin{eqnarray}
w_{\rm scr+shft}=n_a \zeta  e^{-(E_r+
\Delta_{E_{r} })/T}
 \exp \Bigr( {-\pi e^2 Z^2 \sqrt { \mu }\over \sqrt {E_r+\gamma \, T +\Delta_{E_r}} }~\Bigr).  
 \nonumber\\
 \,
\label{ans3}
\end{eqnarray} 
The question now is, ``what is $\Delta{E_r}$?". Both ref. \cite{cussons} and ref. \cite{itoh} choose it to be given by 
$E_{12}-2 E_6$, where $E_Z$ is the free energy shift of an ion with charge $eZ$ due to interaction with the plasma.
In this case, in the ``basically classical " approach it is a theorem that $\Delta_{E_{r}}=V_{sc}(0)$ \cite{alastuey}. But in the next section
we shall argue that the energies of the incoming $^6C$ ions \underline{probably} should not have been subtracted, in which case we would have had 
$\Delta_{ E_r}\approx 2 V_{sc}(0)$. 
What do we mean by ``probably" ? Actually, only that if we were to start from what we believe to be the correct
\it{ab initio} \rm approach, and ask how, making reasonable assumptions, we can get anything like the above structure,
then we would come to the latter conclusion. When we look more deeply into the problem, however, it looks more and more
as though nothing about this whole general approach can be trusted. 

For now,  just in order to show the importance of the issues, we compare the two suggestions made above
for the parameter range specified above.
In fig.1 we compare the ratios of screened to unscreened rates for our two different assumptions as to the value
of $\Delta_{E_r}$, plotted over a range of temperatures.

\vspace{.0 in}
\begin{figure}[h] 
 \centering
 \includegraphics[width=4in]{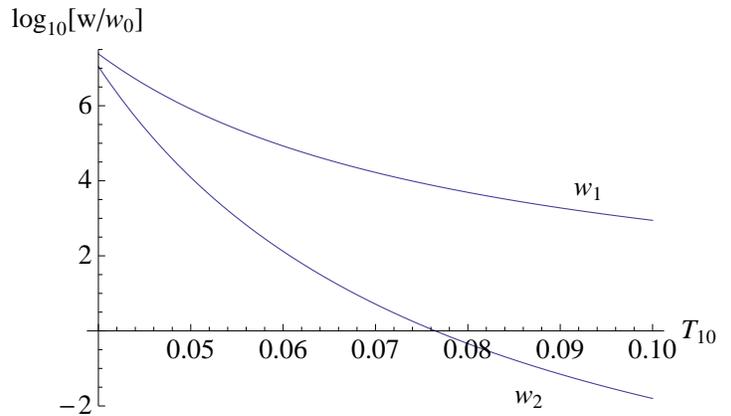} 
  \caption{\small The logarithms of the ratios of the screened rates, $w_1$ and $w_2$ to unscreened rates $w_0$ from (\ref{ans3}).
The rate $w_2$ has been calculated taking $\Delta_ {E_r}$ as just the resonance energy shift, while the rate $w_1$ is the result if, in calculating $\Delta_{E_r}$ we also subtract the energy shifts of the incoming nuclei.  }
  \label{fig:1}
\end{figure}

 Looking at fig. 1, we can stipulate that the subject of this paper can be important, even in a region of moderate
values of the plasma coupling. But we are not here addressing data, nor the likelihood of existence of the possible
resonance. We are only examining the underlying theory, and
we must characterize (\ref{ans3}) as an {\it ansatz}, rather than as a derived result.
This is in contrast to what we would argue {\it vis-a-vis} the case of most applications of (\ref{old})
to non-resonant fusion, but even the latter is subject to the following caveat:
eq. (\ref{old}), giving the rate correction in terms of $K$, is true \underline{only} in the case
that the energy release $Q$ in the fusion is much greater than the Gamow energy, $E_G$ \cite{bs} \cite{rfs1}. For the case of resonant fusion the analogue to the value of $Q$ is, in effect, not only not large compared to $E_G$, it is negative. 
It is our opinion that we can only begin serious consideration of plasma effects in the resonant 
case only after developing a correct formal framework for the discussion. This is taken up in the next section.

 \section{2. Formal framework}

We quote some general relations that are derived in the appendix.
In these derivations for the fusion 
reaction, A+B$\rightarrow$R, we take the
resonance, R, as a particle, with a point coupling to the fusing nuclei. Described in terms of
field operators $\psi_a({\bf r})$ and
$\psi_b({\bf r})$ for the fusing particles, and $\psi_r({\bf r}) $ for the resonance, a zero-range fusion interaction is,
\begin{eqnarray}
H_{\rm fus}=g \int d{\bf r}\,\psi^\dagger_r \,\psi_a \,\psi_b \, +H.C.\,,
\label{hfus}
\end{eqnarray}
where calculation of the decay rate for R$\rightarrow$A+B gives $g^2$ in terms of $\Gamma^{(0)}_r$,
 \begin{eqnarray}
g^2=2^{-1/2} \pi\,\mu^{-3/2}\, \Gamma^{(0)}_r E_{r}^{1/2}\,.
\end{eqnarray}

Since the above model gives back exactly the Breit-Wigner formula when we sum the chain
of repeated interactions to find the scattering amplitude, we judge it to be adequate for our
purposes, as long as the resonance width in the fusion channel is small. The total width
will not enter the explicit results, but for the whole application we assume that the decay
is mostly into other channels. This simulation of the resonance should be good as long as
the barrier penetration calculation doesn't care about the difference between ${ r=0}$ and
the nuclear radius.

For purposes of exposition we think of singled-out reacting particles, A and B. The
``plasma" is composed of all other particles. The rate will then have a factor $V^{-2}$ where $V$
is the volume; when we restore translational invariance and have finite densities of the species A and B
this factor is to be replaced by the product of number densities $n_a n_b$ to get rate per volume, or just
by $n_a$ to get rate per nucleus B as in our convention stated above. We make this replacement 
{\it ab initio} in what follows. It will also eliminate superfluous complication from what follows to take both the nucleus B (described by $\psi_b$)
and the resonance R to have infinite mass, and to position them at the origin. The corrections, expressed as a
multiplicative factor, will be very nearly the same as in the equal mass case.

With this simplification, in the appendix we derive
expressions (\ref{ori}) and(\ref{ans6}), which together give the resonant fusion rate, 
 
 \begin{eqnarray} 
&w=g^2 \,Z_P^{-1}n_a   \int_{-\infty}^\infty dt\,  
\nonumber\\
&\times \sum_{\rm Pl, Pl'}\Big [\Bigr \langle {\bf  r_a=0}, {\rm B}, {\rm Pl} \Big | e^{-(\beta -it) H} \Big |{\bf r_a=0},{\rm B} ,\,{\rm Pl'}\,\Big \rangle 
\nonumber\\
&\times \Big \langle R,\,{\rm Pl'}\Big |  e^{-it  H} \Bigr | R,{\rm Pl}\Bigr \rangle\Big ]\,,
\label{origa}
\end{eqnarray}
where $\beta=1/T$ and $Z_P$
is the partition function for the plasma. We have used the notation, Pl, to stand for all of the coordinates of the plasma particles collectively. The symbol B within the first bracket indicates that nucleus B is present (at $r=0$)
in the space of states in which this calculation is to be carried out. In the second bracket we have only the plasma
and the single nuclear state R, the latter fixed at the origin.

Now we ask: ``What is the relation between (\ref{origa}) and the {\it ansatz} (\ref{ans3}) ?" We begin
by ignoring the coupling of the resonance state $R$ to the plasma. For the final bracket in (\ref{origa})
we have simply,

\begin{eqnarray}
 \Big \langle R,\,{\rm Pl'}\Big |  e^{-it  H} \Bigr | R,{\rm Pl}\Bigr \rangle =e^{-i E_r t} \delta({\rm Pl}-{\rm Pl'})\,,
 \label{finalfactor}
 \end{eqnarray}
where the delta function is the product of delta functions for the individual plasma particles. Then the sum 
over plasma states in (\ref{origa}) gives the dependence $\exp(-i E_r t)$ times $K({\bf 0},t)$, where $K$ is the
correlator as used in (\ref{old}), but now extended to finite time by the replacement $\beta\rightarrow \beta-it$
and in the form of a trace over the plasma coordinates ${\rm Pl}$. In the ``basically classical" approach
it is just,
\begin{eqnarray}
K({\bf 0},t)=\int {d^3 p \over (2 \pi)^3 }e^{-E_p (\beta -i t)}|\Psi_p({\bf 0})|^2\,,
\label{prefactor}
\end{eqnarray}
where the $\Psi_p$ is the wave-function at the origin in the screened potential. For the case considered in the
introduction in which the
screening part changes little within the classically inaccessible region we obtain,
\begin{eqnarray}
|\Psi_p ({\bf 0})|^2\approx 2 \pi e^2 Z^2 |{\bf p} |^{-1}\exp \Big(-\pi e^2 Z^2 \sqrt{2\mu\over E_p-V_{sc}(0)}\,\Big)\,.
\nonumber\\
\,
\label{coul}
\end{eqnarray} 
Putting (\ref{prefactor}), (\ref{finalfactor}) into (\ref{origa}) and using {(\ref{coul}) we obtain (\ref{ans2}), 
later to be modified by the plasma induced energy shift. Next we ask, ``Does the governing
formula (\ref{origa}) tell us how to do this modification"? Note that once we put in interactions with the plasma
the final factor in (\ref{origa}), $\langle R,Pl'|...|R,Pl\rangle$ will not be a delta function in the plasma coordinates, nor diagonal
in these coordinates. Thus we cannot eliminate the sum over states ${\rm Pl'}$. Moreover the first
bracket is no longer a trace and is then no longer given terms of the correlation function. 

However, in the spirit of ``basically classical", we can say ``We determined the effective potential
between ions A and B as the logarithm of the classical correlation function, in a calculation in which traced over
the plasma states. So we now imagine that when A and B are close
to each other the plasma is in a single configuration, ${\rm Pl_0}$, a configuration that produces an electrical potential $\phi_{\rm sc}$ at the origin. The sum of the interaction energies
of these two particles with this potential, less the Coulomb energies of the individual widely separated ions, gives the screening
potential energy $V_{\rm sc}(0)$."

With this state ${\rm Pl_0}$ replacing both ${\rm Pl}$ and ${\rm Pl'}$ in (\ref{origa}), and eliminating the sum over Pl',
the second bracket then becomes,

\begin{eqnarray}
 \Big \langle R,\,{\rm Pl_0}\Big |  e^{-it  H} \Bigr | R,{\rm Pl_0}\Bigr \rangle =e^{-i (E_r+\Delta{_{E_r}})t }\,,
 \label{finalfactor2}
 \end{eqnarray}
which leads back to (\ref{ans3}). 

With this pseudo-derivation in hand, we feel more confident in declaring our belief that \underline{if} any approach
that leads to the form (\ref{ans3}) makes sense, for strongly coupled systems, then
 $\Delta_{E_{r}}$  \underline{must} be simply the energy shift of the resonance R in the plasma, as
given by $e^2(Z_1+Z_2) \phi_{sc}$, with
 \underline{no} subtractions for the energy shifts of the isolated A and B states in the same plasma.
 Indeed, the object estimated in (\ref{finalfactor2}) has no way of knowing how the total charge $Z_1+Z_2$
 was split between the incoming nuclei.
 The above conclusion is at variance with the assumptions of refs. \cite{cussons} and \cite{itoh}, and we have 
 already illustrated in fig. 1 the very big difference that the choice makes in the rate formula.
 
 That said, we believe that it is much more likely that the basically classical approach, which appears
 to have served us well in ordinary fusion reactions, when the energy release $Q$ is much greater
 than the Gamow energy,
 simply cannot be used to get meaningful results in this resonance fusion problem, with any choice of $\Delta_{E_r}$.

We best can explain the essential difference between the resonance case and the normal fusion case by recalling
some of the features of the latter. Again, strictly for economy of exposition, in the reaction ${\rm A+B\rightarrow C+D}$,
we take the two ions B and D to have infinite mass. As in the resonance case we take a zero range fusion interaction.
Instead of a resonance energy parameter we now have an ordinary energy release parameter, $Q$. The analogue to
(\ref{origa}) is now, as given in ref. \cite{rfs1},

\begin{eqnarray} 
&w=g^2 Z_P^{-1}\, n_a  \int_{-\infty}^\infty  dt \, e^{i Q t} 
\nonumber\\
&\times \sum_{\rm Pl, Pl'}\Big [
\Bigr \langle {\bf r_a=0}, {\rm B}, {\rm Pl} \Big | e^{-(\beta -it) H} \Big |{\bf r_a=0},{\rm B} ,\,{\rm Pl'}\,\Big \rangle 
\nonumber\\
&\times \Big \langle {\bf r_c=0}, {\rm D},\,{\rm Pl'}\Big |  e^{-it  H} \Bigr |{\bf r_c=0}, {\rm D},{\rm Pl}\Bigr \rangle\Big]\,.
\label{origb}
\end{eqnarray}

Now we can see how in the ordinary fusion case, when $Q>>E_G$ the ``basically classical" approach may suffice.
In the A+B $\rightarrow$C+D fusion process of (\ref{origb}) when
the energy release $Q$ is large compared with the Gamow energy, it is a reasonable approximation to set $t=0$ in the exponent
within the first bracket $\langle\rangle$ (pertaining to the initial A + B system). If one further omits all plasma coupling
of the fusion products to the plasma in calculation of the second bracket (pertaining to the C+D system), and keeps just the kinetic energy term, $H_0$, then the time integral
just sets the energy of the fusion products to the value $Q$. Since we have taken the operator 
in the second bracket to be independent of the plasma coordinates, we can integrate $\delta ({\rm Pl-Pl' })$
over the intermediate plasma configurations ${\rm Pl'}$, and the first $\langle\rangle$, integrated over the space Pl, becomes the trace defining the static correlator, bringing us back to the standard result (\ref{old}). 

But for the resonance case we find no similar argument that we can make to defend (\ref{finalfactor2}) 
or to lead back to (\ref{ans3}).

\section{3. Perturbative results}
As an alternative to the  machinations of the last section, we can, beginning from (\ref{origa}), turn the perturbation crank to generate results for the weak coupling case. Though the results would of course not be applicable to the strongly
coupled systems discussed earlier, we could hope that they would cleanly resolve the question of what to use
for the resonance energy shift in (\ref{ans3}), assuming for the moment that the calculation leads to the
general structure of (\ref{ans3}). But it will not, as we shall see. Nonetheless, in the appendix we do
work out these corrections in detail. For one thing, they comprise the only certain result as to the plasma
corrections in the resonant case.
For another, they highlight problems that were swept under the rug in the discussion that led to
(\ref{finalfactor2}), and that raise doubts about any efforts to make a connection to the ``basically classical"
lore, as used in our speculations above. 

First we give the result. The expansion is in powers of the couplings of the plasma particles to each other and to the fusing particles. The $e$'s in the Coulomb
wave-functions of the fusing particles are not part of the expansion, of course. Because of the long range
of the Coulomb potential, the leading corrections are of order $e^2\kappa_D$,
where $\kappa_D$ is the Debye momentum. 

After a long calculation, given in the appendix, involving the cancellation of many terms of this order, we find simply,
\begin{eqnarray}
 w=[1+e^2Z_1 Z_2\kappa_D \beta]w_0\,,
 \label{salp}
\end{eqnarray}
 which is just the original Salpeter correction, as used in standard weak-coupling applications, most particularly
in the sun since that is the only venue in which relatively small corrections 
are of much interest. Is this what we did so much work to learn, that the effect of the plasma is still
just in the initial state equilibrium contact probability? That is, whether the final system is resonant or not,
we have the same modification factor? Indeed if we look back at the \it ansatz \rm, (\ref{ans2})
and choose $\Delta_{E_r}=-\gamma T$, as in the case in which the screening potential does not
change appreciably within the classically inaccessible region, we recover the equivalent result, namely just
the same screening factor $\exp (\gamma)$ that would modify a non-resonant fusion \footnote{This was pointed
out in ref. \cite{cussons}.}.

However, in the perturbative calculation in the appendix we see that the simple result depends completely
on including the terms which, in the notation of (\ref{origa}), the intermediate plasma states ${\rm Pl'}$ are
different from the initial plasma states in the trace, Pl.
Described in terms of fundamental processes we can say that the incoming nuclei scatter a plasma particle
and then this plasma particle is put back into its initial state through an interaction with the final resonance;
or in an equivalent description in terms of scattering amplitudes rather than correlation functions,
that we must calculate interference terms between amplitudes in which a plasma particle scatters from a fusing nucleus
and amplitudes in which it scatters from the resonance.

If we left out terms like this and calculated only the separate contributions from the interaction of the plasma
with the initial state and from the interaction with the resonance, then the sum of these terms would not be given
by (\ref{salp}). Depending on the parameters, it can be an order of magnitude greater than shown in (\ref{salp}).
Yet going back to the strong coupling case, combining the two effects, classical screening and energy shift,
is exactly what was prescribed. So it appears to us that it is not meaningful to calculate
this way, and that, for example, neither of the ``energy shift" models that we used in the comparison
of fig. 1 are viable.

For comparison, if we did the same kind of perturbation calculation for the normal fusion case
we would again obtain the Salpeter result with additional terms that cancel when we add them up.
In contrast to the resonance case, however, these individual terms are all reduced by at least one order
of the factors $(Z_1Z_2 e^2 \kappa_D)/Q$ or $E_{\rm Gamow}/Q$, as long as these ratios are small. 
Thus in this case the perturbation analysis gives us no cause to question the qualitative argument for
the ``basically classical" approach given at
the end of the last section.
In ref. \cite{rfs1} we have argued the need for rethinking the plasma corrections for cases in which $Q$ is so
small that the condition is not fulfilled.

\section{4. Discussion}

Our strongest conclusion in this paper, which lacks definitive computational results for the systems
of most interest, is that one should begin with a complete and correct formulation. It appears to us that
the literature on this topic has depended on using physical intuition to
graft together some pieces of lore that have served well in
other contexts (but with some limitations that are usually unacknowledged).
We do not believe that this provides
a coherent approach starting from basic physics.

We list the two new results of this paper that have been proven, then turn to speculation:

1). The formula for reaction rate, (\ref{origa}). This was stated for the case in which 
one of the fusing particles and the final resonance both have very large mass, so that they each
would stay at the origin. But it provides an adequate testing ground
for the general questions that we raised in relation, e.g., to $^{12}C+^{12}C$ fusion.
\footnote{
The corresponding formula for the case of general masses is easy to derive, but more
complex. In the bracket for the fusing particles, the two particles that meet at ${0}$ in the
left hand state,$\langle~|$, meet again in the right hand state, $|~\rangle$, but now at position ${\bf r}$.
Correspondingly, in the bracket for the fusion product R we go from position ${\bf r}$
to position {\bf 0}. The whole expression has an additional integral over ${\bf r}$; all of
this serves to transfer total momentum from the initial to the final state. Center of
mass motion is no longer totally trivial when we are in a plasma and $Q$ is insufficiently
large. The derivation of the above follows easily from the methods used in the
appendix. For the ordinary fusion case the analogue is shown in detail in ref. \cite{bs}, with
a somewhat different representation of the operators involved, however.}

2). The summation of the leading correction terms in the perturbative development. The outcome
argued against our favorite guess as to how to do the energy shift calculation, but the details did show the necessity
of following the plasma in a way in which one does not in the ``basically classical" approach. We characterize
the procedure of the latter as,
 ``Do one big classical simulation, for the effective potential, and thereafter do your fusion calculation
with the individual plasma states out of the picture." There is evidence \cite{militzer} that this procedure
is adequate for determining the contact probability in the initial state, for an ordinary fusion reaction, as long as
$E_{\rm Gamow}<<Q$.
But when $Q$ is small, or we are dealing instead with resonance fusion, we believe that it is insupportable.

If we had to choose between the two alternatives offered in the first section,
we would follow the discussion just before and after (\ref{finalfactor2}) and use the energy shift
of the resonance particle itself for $\Delta_{E_r}$, with no subtractions for the energy shifts of
the incoming particles. This gives the large relative suppressions that we see in the $\log[w_2/w_0]$ curve in
fig. 1.

Although in the above opinion we have reverted to the kind of guesswork that we deplored in the introduction,
we do add that the perturbative results do not argue as strongly as may have appeared for using the other option for $\Delta_{E_r}$, as adopted
by refs.\cite{cussons} and \cite{itoh}.
First, suppose that we separate out the the normal Salpeter factor, $(1+e^2Z^2\kappa_D \beta)$, from all other
terms. Now the \underline{sum} of these other terms is convergent even if we calculate them with unscreened Coulomb
interactions, i e., the infrared divergences in the individual terms have cancelled against each other. 

This behavior is similar in form to the cancellation of (photon) infrared divergences in the inclusive
cross-sections for charged particle scattering, where individual contributions, like photon bremsstrahlung,
are divergent in an order by order expansion. The point we make is that these cancellations do not
carry forward in any fashion to the non-infrared-divergent parts. Returning back to our perturbative
calculation, we believe that the cancellations of the infrared terms that restored the Salpeter form
do not extend in any sense to the non-infrared divergent terms (which begin with terms of order $e^4$ but
also include higher corrections of order $\Gamma^3$). 
Therefore, for applications in which the coupling is even of moderate strength, the only thing
that we really learn from the elaborate calculations in the appendix is that it is wrong to
disregard the changes in the plasma, i. e. the sum over the states $|{\rm Pl'}\rangle$
in (\ref{origa}). 

 We ask how one might imagine doing a real calculation, assuming unlimited computing resources.
The guiding example could be the calculations of Militzer and Pollack \cite{militzer} or Ogata \cite{ogata}, who do real
quantum path integral evaluations for the static correlator at small distances, without the
assumptions of the ``mostly classical" approach. In ref \cite{militzer} the authors write $e^{-\beta H}=[e^{-\beta H/N}]^N$
and choose $N$ sufficiently large so that each factor $\langle {\rm Pl'}| e^{-\beta H/N}|{\rm Pl''}|\rangle$,
effectively at high temperature, can be expanded in perturbation theory. The product of the $N$ factors,
involving N integrals over intermediate plasma configurations, is very expensive to compute, needless to say, but because of
the negative real exponentials the calculation is feasible. In contrast, the result (\ref{origb}), with its factor $e^{-(\beta-it)H}$
and extra time integral, cannot be calculated directly by such a procedure, because of the oscillations. Clearly some
technical progress is greatly needed.

An important fusion chain that requires resonance enhancement is the triple $\alpha$ process. There
are suggestions for treating the associated plasma effects in ref. \cite{lamb}. In red giant cores where the process
makes its first appearance, the plasma coupling is too weak to make very much difference. But when, e.g., 
it takes place in surface layers of accreting neutron stars, all of the caveats of the present paper apply. 

We also note recent considerations of the role of a possible $^12$C+$^12$C resonance on the ignition of a
type 1a supernovae \cite{CC1} \cite{CC2}. Though screening issues were not addressed in either of these works,
the plasma coupling is strong enough in this application that they need to be. \footnote{Our refs. \cite{cussons} and \cite{itoh}
are cited in \cite{CC1}, but their results not used in the analysis.}

This work was supported 
in part by NSF grant PHY-0455918. 
\begin{widetext}
\appendix*
\section{Appendix}
\setcounter{equation}{0}

For completeness we begin by recapitulating some development from ref \cite{bs}. 
We calculate the time rate of change of the species A induced by the fusion interaction (\ref{hfus}). Since the system is translationally invariant we can choose to evaluate this time derivative at 
at point ${\bf r}={\bf 0}$ and choose time to be zero as well. Directly from the Heisenberg equations and the commutation
rules, writing
the transform of (\ref{hfus}) as,
\begin{eqnarray}
H_{\rm fus}(t)=\int d{\bf r}\,\, [h({\bf r},t)+h({\bf r},t)^\dagger]\,,
\end{eqnarray}
where $h=g\psi_a \psi_b\psi_r^\dagger$ we obtain the rate of change of $n_a$, the density of species A,
\begin{eqnarray}
\Bigr \langle \stackrel{\cdot} n_a({\bf 0,0})\Bigr \rangle_\beta=- i\langle [n_a({\bf 0},0),H_{\rm fus} (0)]\Bigr \rangle_\beta=i\Bigr \langle[h({\bf 0},0)-h^\dagger({\bf 0},0)]\Bigr \rangle_\beta.
\label{ndot}
\end{eqnarray}

The notation $\langle...\rangle_\beta$ indicates the thermal average in the medium, such that for an operator, $O$, we have
 \begin{eqnarray}
 \langle O\rangle_\beta\equiv Z_P^{-1}{\rm Tr} \Big [O\exp(-\beta [H+H_{\rm fus}- \mu_a N_a-\mu_b N_b])\Big]\,,
 \end{eqnarray}
where $Z_P$ is the partition function, and where $H$ is the Hamitonian in the absence of the fusion term. We wish to calculate the rate to
lowest non-vanishing order; i.e. to second order in $h$.  Thus we now must consider the linear response of the average of the operator $[h({\bf 0},0)-h^\dagger({\bf 0},0)]$, as it appears in (\ref{ndot}), to the perturbation $H_{\rm fus}$, giving an expression for the rate,
\begin{eqnarray}
w=-i \int_{-\infty}^0 dt \int (d{\bf r})\Bigr \langle [  \stackrel{\cdot} n_e({\bf }0,0),H_{\rm fus}(t)]\Bigr \rangle_\beta
=\int_{-\infty}^0 dt \int d{\bf r}\Bigr \langle [h({\bf 0},0)-h^\dagger({\bf 0},0)],[h({\bf r},t)+h^\dagger({\bf r},t)]\Bigr \rangle_\beta 
\nonumber\\
\,
\label{w}\,,
\end{eqnarray}
where now the thermal average in the medium is to be calculated under the action of $H$ alone. 
We note that,
\begin{eqnarray}
\Bigr\langle [h({\bf 0},0), h({\bf r},t)]\Bigr \rangle_\beta=
\Bigr\langle [h^\dagger ({\bf 0},0), h^\dagger({\bf r},t)]\Bigr \rangle_\beta=0\,.
\label{hcom}
\end{eqnarray}
Using in addition the space-time translational invariance of the medium and the antisymmetry of
the commutator,
\begin{eqnarray}
\Bigr\langle [h^\dagger ({\bf 0},0), h({\bf r},t)]\Bigr \rangle_\beta=-\Bigr\langle [h ({\bf 0},0), h^\dagger({\bf -r},-t)]\Bigr \rangle_\beta
\,,
\end{eqnarray}
we can write the rate as
\begin{eqnarray}
w
=-\int_{-\infty}^\infty dt \int d{\bf r}\Bigr \langle [h({\bf 0},0),h^\dagger({\bf r},t)]\Bigr \rangle_\beta \,.
\label{w2}
\end{eqnarray}

When we take the medium to contain no nuclei of type R, so that there is no reverse reaction, we can omit the
first term in the commutator in (\ref{w2}). Then inserting (\ref{hfus}) we obtain,
\begin{eqnarray}
 w=g^2  \int_{-\infty}^\infty dt \int d{\bf r}\,e^{-iE_r t}\Bigr\langle \psi_a^\dagger({\bf r},t)\psi_b^\dagger({\bf r},t)
 \psi_r({\bf r},t)\psi_r^\dagger ({\bf 0},0) 
  \psi_a({\bf 0},0)\psi_b({\bf 0},0)
\Bigr \rangle_\beta
\equiv g^2 \int_{-\infty}^{\infty} dt ~y(t),
\label{ori}
\end{eqnarray}
where,
\begin{eqnarray} 
y(t)= Z_P^{-1} \int d{\bf r}{\rm Tr}\Bigr[ \Lambda e^{-(\beta -it) H}\psi_a^\dagger({\bf r},0)\psi_b^\dagger({\bf r},0) \psi_r({\bf r},0)e^{-it  H} \psi_r^\dagger ({\bf 0},0)\psi_a({\bf 0},0) \psi_b ({\bf 0},0) ]\,,
\label{orig}
\end{eqnarray}
and where $\Lambda=\exp[\beta \sum \mu_i N_i]$.

The relation (\ref{ori}) is the fundamental formal result
required to determine the fusion rate. We have taken the Schrodinger and Heisenberg representations to coincide
at $t=0$, and at this time the $\psi$ operators can be expanded in annihilation operators for single particle
states.

The calculation that follows will be much simpler if we specialize to the case in which one of the fusing particles, say B, and
the resonance, R, are much heavier than ion A; furthermore the plasma corrections to the order that we calculate, as a multiplicative factor,
do not depend at all on the mass configuration. Therefore we simplify, {\it ab initio }, by taking the masses of B and R to be infinite,
and situating a single B at $\bf r=0$. We replace operators $\psi_b({\bf r},t) , \psi_r({\bf r,t})$  by operators $b(t)$, and $r(t)$ that
respectively annihilate the indicated particles situated at the origin. Note that even though ions B and R no longer carry kinetic energy, these Heisenberg operators are still time dependent by virtue of the Coulomb interactions of the ions to which
they refer. The result (\ref{orig}) is now replaced by ,
\begin{eqnarray} 
y(t)= Z_P^{-1} V^{-1} {\rm Tr }\Bigr[\Lambda  e^{-(\beta -it) H}\psi_a^\dagger ({\bf 0},0) b^\dagger (0) r(0)e^{-it  H} r^\dagger (0)  
\psi_a({\bf 0},0) b(0) \Bigr ]\,.
\label{ans5}
\end{eqnarray}

The rate that we calculate is now the rate of production of the resonance from the single ion B situated at ${\bf r=0}$. When we 
recast the expression in a form in which we consider only one A particle, the factor $V^{-1}$ in (\ref{ans5}) gets replaced by a factor
of $n_a$. Written a bit more explicitly to exhibit the plasma coordinates in an intermediate state (\ref{ans5}) reads,
 \begin{eqnarray} 
y(t)=Z^{-1} n_a
\times \sum_{\rm Pl, Pl'}\Big [\Bigr \langle {\bf r_a=0}, {\rm B}, {\rm Pl} \Big | e^{-(\beta -it) H} \Big |{\bf r_a=0},{\rm B} ,\,{\rm Pl'}\,\Big \rangle 
 \Big \langle R,\,{\rm Pl'}\Big |  e^{-it  H} \Bigr | R,{\rm Pl}\Bigr \rangle\Big ]\,,
\label{ans6}
\end{eqnarray}
which leads directly to (\ref{origa}).
 
To calculate the leading terms in a perturbation expansion we write the
total Hamiltonian (not including $H_{\rm fus}$) as $H=H_0+H_I$ , where $H_0$ includes all kinetic energies as well as
the mutual Coulomb interactions of the plasma particles with each other and the Coulomb interaction between the distinguished nuclei A and B. Then $H_I$ simply contains the Coulomb couplings between the nuclei, A , B, C and the 
plasma particles. Designating the total electrical potential of the plasma particles at position
$\bf r$  as $\phi({\bf r})$ we have,
\begin{eqnarray}
H_I=\int d{\bf r}~ e_a n_a({\bf  r})\phi({\bf r})+[\,e_b \,b^\dagger\,b+e_r\,r^\dagger \,r\,]\phi({\bf 0})\,.
\label{HI}
\end{eqnarray}

Of course, if we were attempting to go beyond the weak coupling regime, we would have to face the fact that the potentials
$\phi$ in (\ref{HI}) involve all of the coordinates of the individual plasma particles and that a numerical simulation,
as will be required for a definitive answer in the strongly coupled case, will involve the explicit taking into account of the paths
of all of these particles. In the discussion section we describe the difficulties in principle of carrying out such calculations.
But in the present paper we calculate to second order in $H_I$, and in addition single out only the terms of order $e^3$
in which the superficial perturbative order $e^4$ has been lowered through the long range of the Coulomb force.  
For this purpose all we need to know about the plasma is the part of the field-field correlation function in momentum space that is singular for small $k$ as $\kappa_D$ goes to zero, namely \cite{bs},
\begin{eqnarray}
\Bigr \langle \phi ({\bf k},t') \phi ({\bf -k'},t'') \Bigr \rangle_{IR}\approx 4 \pi \delta ({\bf k-k'})
\beta^{-1}\kappa_D^2{1\over (k^2+\kappa_D^2)k^2}
\label{corr}\,.
\end{eqnarray}
We introduce the interaction picture through the identities,

\begin{eqnarray}
e^{-H_a (\beta-it)}=e^{-H_0(\beta-i t)} \Omega_+(-i\beta, t)~~~~~~~;~~~~~
e^{-i H_b t}= \Omega_-(0,t) e^{-i H_0 t }\,,
\label{int}
\end{eqnarray}
where,
\begin{eqnarray}
\Omega^{(+)}(-i\beta, t)=\exp\Bigr [i \int_{-i \beta}^t dt' \hat H_I(t')\Bigr ]_+ ~~~~~;~~~~~~
 \Omega^{(-)}(0,t)=\exp\Bigr [-i \int_{0}^t dt' \hat H_I'(t')\Bigr ]_- \,,
\end{eqnarray}
time-ordered and anti-time ordered, respectively, where,
\begin{eqnarray}
\hat H_I(t)=e^{i H_0 (t+i\beta)} H_I  e^{-i H_0 (t+i\beta)} ~~~~~~~; ~~~~~~~~
\hat H_I'(t)=e^{-i H_0 t} H_I e^{i H_0 t}\,.
\end{eqnarray}
Substituting in (\ref {orig}) we obtain,
\begin{eqnarray} 
y(t)= Z_P^{-1}V^{-1}{\rm Tr} \Bigr [\Lambda \psi_a({\bf 0},0) b (0) e^{-(\beta -it) H_0}\Omega^{(+)}
(-i\beta, t)
 \psi_a^\dagger ({\bf 0},0) b^\dagger (0) r(0)\Omega^{(-)}(0,t) e^{-it  H_0} r^\dagger ({\bf 0},0)
   \Bigr]\,,
\label{ans7}
\end{eqnarray}
In (\ref{ans7}) we have also used the cyclic property of the trace to move the operators $ \psi_a({\bf 0},0) b (0)$ to
the front; in this form the needed diagonal matrix elements in the trace are between states with no B or R present.
We now expand the $\Omega$ factors in powers of the coupling, retaining only terms of order $e^2$, (or $e_a^2, e_a e_b$, etc.). Schematically, each will have powers of $e$ coming from the thermal expectation of a product of two
ion-electric-potential $\phi({\bf r})$ operators; these are determined by the Hamiltonian $H_0$. If this potential-potential
correlator is itself expanded in powers of $e$, the expansion begins with terms of order $e^2$, so that the
rate corrections superficially would be of order $e^4$ but the small $k$ singularity of (\ref{corr}) reduces the order to $e^3$. In the present paper we
pursue only these $e^3$ terms; the neglected terms will go as power $e^4$ and higher.

Expanding, we obtain,
\begin{eqnarray}
\Omega ^{(+)}(-i\beta,t)=1-i \int_{-i\beta}^t dt_1 \hat H_I(t_1)-\int_{-i\beta}^t dt_1 \hat H_I(t_1)\int_{-i\beta}^{t_1}dt_2 
\hat H_I(t_2)\,,
\label{oma}
\end{eqnarray}
and
\begin{eqnarray}
\Omega ^{(-)}(0,t)=1+i \int_{0}^t dt_1 \hat H_I(t_1)-\int_{0}^t dt_1\hat H_I(t_1)\int_{0}^{t_1}dt_2 \hat H_I(t_2)\, .
\label{omb}
\end{eqnarray}

Our task is to put (\ref{oma}) and (\ref{omb}) into the rate equation (\ref{ans7}) and select out the terms that are of order
$e^2 \kappa_D$. We define $y(t)=y_1(t)+y_2(t)+y_3(t)$ where $y_1,y_2,y_3$ come respectively from the second
order piece of $\Omega ^{(+)}(-i\beta,t)$; the product of the first order parts of $\Omega ^{(+)}(-i\beta,t)$ and
$\Omega_b^{(-)}(0,t)$; and the second order part of $\Omega_b^{(-)}(0,t)$. The operator under the trace now factors into
a part with a  $\phi ({\bf k},t') \phi ({\bf -k'},t'')$ and a part that depends only on the operators in the space of the 
nuclei A,B,R. To the order that we need to calculate, the partition function also factorizes into a product in each space.
Therfore we can directly use (\ref{corr}) to evaluate the plasma part of the trace. Since the electric potential
of the plasma, $\phi$ occurs in $H_I$ in the term $\int d{\bf k} \,n_a({\bf k})\phi({\bf k})$ the ${\bf k}$'s in the
in (\ref{corr}) in the plasma factor still link to the ${\bf k}$'s in the fusing-particle factor. However since the leading
(order $e^3$) behavior comes from very small k, we can evaluate at $k=0$ in the fusing particle sector
and use,

\begin{eqnarray}
\int d{\bf k}\,d{\bf k'} \Bigr \langle \phi ({\bf k},t') \phi ({\bf -k'},t'') \Bigr \rangle_{IR}= (2\pi)^3\beta ^{-1}\kappa_D\,.
\end{eqnarray}

Furthermore we can make the replacement, for example,  

\begin{eqnarray}
\hat H_I=e_a e^{i H_0 t}\int {d{\bf p} \,d{\bf k}\over (2\pi)^6}\, a^\dagger _{\bf p} a_{\bf p+k} \phi({\bf k})e^{-i H_0 t}
\rightarrow e_a \int {d{\bf p}\over (2 \pi)^6} \, a^\dagger _{\bf p} a_{\bf p+k}\int d{\bf k}\, \phi(\bf k,t)\,,
\nonumber\\
\,
\end{eqnarray}
since in the limit of small ${\bf k}$, $H_0$ commutes with $\int d{\bf p}\, a^\dagger_{\bf p} a_{\bf p+k}$.
Then we see that in the infrared limit there is effectively no $t_1,t_2$
dependence of the integrands in \ref{oma}) and (\ref{omb}); performing these integrals gives the respective factors,
\begin{eqnarray}
f_1(t)={\beta^2 \over 2}-{t^2\over 2}-i \beta t\,,
\nonumber\\
f_2(t)=(t+i \beta)t\,,
\end{eqnarray}
and
\begin{eqnarray}
f_3(t)=-{t^2\over 2}\,.
\end{eqnarray}
With the $t$ dependent factors separated from the matrix elements of operator products that we still have to evaluate we 
now write, 
\begin{eqnarray} 
&y_1(t)=\beta \kappa_D f_1(t)Z_a^{-1}V^{-1} \Big \langle 0_b,0_r \Big |{\rm Tr_a}\Lambda_a\Bigr \{ \Bigr [\sum_{ {\bf p}}a_{{\bf p}} 
\Psi_{\bf p} ({\bf 0})e^{-(\beta -it)E_{\bf p} }\Bigr] b \nonumber\\
&\times  \Bigr [e_a \,\sum_{\bf q}a^\dagger_{\bf q} a_{\bf q}
+e_b b^\dagger b\Bigr]
\Bigr [e_a \,\sum_{\bf q'}a^\dagger_{\bf q'} a_{\bf q'}
+e_b b^\dagger b\Bigr] \Bigr[\sum_{{\bf p'}}a^\dagger _{\bf p'}\Psi_{\bf p'}({\bf 0})^*\Bigr]\, b^\dagger r r^\dagger \, \Bigr\}
\Big |0_b,0_r \Big \rangle
\nonumber\\
&=(e_a+e_b)^2 f_1(t)W(t)\,,
\nonumber\\
\,
\label{y1}
\end{eqnarray}
where $\Big |0_b,0_r \Big \rangle$ stands for the state in the B,R part of the space that contains nothing;
$\Lambda_a=\exp(\beta \mu_a N_a)$; the trace acts in the A+plasma space; and,
\begin{eqnarray}
W(t)=2^{-3/2} \pi^{-2}M_a^{3/2}\beta \kappa_D  V^{-1} e^{\mu_a\beta}
\int dE\,\sqrt{ E}\,e^{-(\beta -it)E }\, \Big | \Psi_E({\bf 0})\Big|^2\,.
\end{eqnarray}
In similar fashion we obtain
\begin{eqnarray}
&y_2(t)=\beta \kappa_D f_2(t) V^{-1} Z_a^{-1}  \Big \langle 0_b,0_r \Big |{\rm Tr_a}\Lambda_a\Bigr \{ \Bigr [\sum_{ \bf p}a_{\bf p} \Psi_{\bf p} ({\bf 0})e^{-(\beta -it)E_{\bf p }}\Bigr] b\Bigr [e_a\,\sum_{\bf q}a^\dagger_{\bf q} a_{\bf q}
+e_b b^\dagger b\Bigr]
\nonumber\\
&\times r \,e_r \,r^\dagger c \,\Bigr[\sum_{\bf p'}a^\dagger _{\bf p'}\Psi_{\bf p'}({\bf 0})^*\Bigr]\,  b^\dagger r^\dagger\,\Bigr \}
\Big |0_b,0_r \Big \rangle\nonumber\\
&=(e_a+e_b)e_c f_2(t) W(t)\,,
\nonumber\\
\,
\label{y2}
\end{eqnarray}
and,

\begin{eqnarray}
&y_3(t)=\beta \kappa_D f_2(t)V^{-1} Z_a^{-1}  \Big \langle 0_b,0_r \Big |{\rm Tr_a}\Lambda_a\Bigr \{[\sum_{ \bf p}a_{\bf p} \Psi_{\bf p} ({\bf 0})e^{-(\beta -it)E_{\bf p} }\Bigr] b
\nonumber\\
&\times r \,e_c^2 \,[r^\dagger r]^2 \Bigr[\sum_{\bf p'}a^\dagger _{\bf p'}\Psi_{\bf p'}({\bf 0})^*\Bigr]\,  b^\dagger r^\dagger\,\Bigr\}
\Big |0_b,0_r \Big \rangle\nonumber\\
&=e_r^2 f_3(t) W(t)\,.
\label{y3}
\end{eqnarray}

In the above we did not impose charge conservation $e_r=e_a+e_b$, in order to best exhibit which interactions
were contributing to which terms in the answer. Now eliminating $e_r$ and summing
the above three pieces, using $f_1(t)+f_2(t)+f_3(t)=\beta^2/2$, we obtain simply,

\begin{eqnarray}
y(t)=
 2^{-5/2} \pi^{-1}M_a^{3/2}\beta \kappa_D  V^{-1}(e_a+e_b)^2 e^{\mu_1\beta}
\int dE\,\sqrt E\,e^{-(\beta -it)E }\, a^\dagger_E a_E \Big | \Psi_E({\bf 0})\Big|^2\,.
\nonumber\\
\,
\end{eqnarray}
Substituting in (\ref{ori}), $w=g^2 \int dt\, y(t) \exp (-i E_r t)$, performing the time integration, then using,
\begin{eqnarray}
g^2=2^{-1/2} \pi^{-1}M^{-3/2} \Gamma_r E_{r}^{1/2}\,,
\end{eqnarray}
and,
\begin{eqnarray}
e^{\beta \mu_a }=n_a (2\pi)^{3/2}(M_a T)^{-3/2}\,,
\end{eqnarray}
and replacing the factor $V^{-1}$ by $n_b$,
we obtain for the change of the fusion rate due to the plasma interactions,
\begin{eqnarray}
\delta w={1\over 2}(e_a+e_b)^2\beta \kappa_D w_0\,,
\end{eqnarray}
where
\begin{eqnarray}
 w_0=n_a n_b \Bigr ( {2 \pi \over M_a T}\Bigr )^{3/2}\Gamma_r e^{-E_r/T} \exp \Bigr( -\pi e^2 Z^2 \sqrt { {M_a \over E_r}} ~~\Bigr)  \,.
\end{eqnarray}
This gives, after the zero'th order term is added,
\begin{eqnarray}
w=w_0[1+{1 \over 2} \beta \kappa_D (e_a+e_b)^2]\,.
\label{semifinal}
\end{eqnarray}
Before drawing conclusions from (\ref{semifinal}) we note that the factor $n_a$ in the zero'th
order term emerged computationally from,
\begin{eqnarray}
n_a=(V Z)^{-1}\sum_{\bf p} \langle e^{-\beta( H-\mu_a N_a)} a^\dagger_{\bf p} a_{\bf p}\rangle =e^{\beta \mu_a} \int {d^3 p \over (2 \pi)^3} 
e^{-\beta p^2 /2M},    
\end{eqnarray}

But if we set out to calculate $n_a$ in our basic formalism we would have had begun instead with,
\begin{eqnarray}
n_a=(V Z)^{-1}\sum_{\bf p} \langle e^{-\beta( H-\mu_a N_a)} a^\dagger_{\bf p} a_{\bf p}\rangle    
= \sum_{\bf p} \langle e^{-\beta H_0}a^\dagger_{\bf p} a_{\bf p} \Omega^{(+)}(-i \beta,0)\rangle 
\nonumber\\
=(1+{e_a^2\over2} \beta \kappa_D) e^{\beta \mu_a)} \int {d^3 p \over (2 \pi)^3} e^{-\beta p^2 /2M} \,, 
\end{eqnarray}
and similarly for $n_b$.
That is to say, the plasma interactions have changed the relation between number density
and chemical potential in just such a way that when the relation (\ref{semifinal}) is rewritten in terms
of the actual density, the terms of order $e_a^2$ (and $e_b^2$) are removed \cite{bs}, leaving us with the final answer 
\footnote{This last argument is given in more detail in ref. \cite{bs} , appendix B},
\begin{eqnarray}
w=w_0[1+e_a e_b \beta \kappa_D ]\,,
\label{end}
\end{eqnarray}
the result (\ref{salp}) quoted in text.

\end{widetext}

\end{document}